\documentstyle[prl,aps,epsfig,amssymb,multicol]{revtex}
\begin{document}
 \draft
\title{Binomial level densities}
\author{ A. P. Zuker}
\address{ IRES, B\^at27, IN2P3-CNRS/Universit\'e Louis
Pasteur BP 28, F-67037 Strasbourg Cedex 2, France}
\date{\today}
\maketitle
\begin{abstract}
  It is shown that nuclear level densities in a finite space are
  described by a continuous binomial function, determined by the first
  three moments of the Hamiltonian, and the dimensionality of the
  underlying vector space. Experimental values for $^{55}$Mn,
  $^{56}$Fe, and $^{60}$Ni are very well reproduced by the binomial
  form, which turns out to be almost perfectly approximated by Bethe's
  formula with backshift. A proof is given that binomial densities
  reproduce the low moments of Hamiltonians of any rank: A strong form
  of the famous central limit result of Mon and French. Conditions
  under which the proof may be extended to the full spectrum are
  examined.
\end{abstract}

\pacs{21.10.Ma,21.60.Cs,21.60.-n,27.40.+z}

\begin{multicols}{2}

\narrowtext

The vast majority of studies on level densities rely on Bethe's
formula\cite{bet36},
\begin{equation}
  \label{bet}
  \rho_B(E,a,\Delta)=
\frac{\sqrt{2\pi}}{12}\frac{e^{\sqrt{4a(E+\Delta)}}} {(4a)^{1/4}(E+\Delta)^{5/4}}.
\end{equation}

Although only two parameters are involved for a given nucleus---and a
third, the spin cutt-off, for work at fixed angular momentum
$J$---their dependence on mass number, excitation energy and shell
effects makes their determination delicate (see \cite{thi97,mug98} for
recent work and earlier references). The problem is that the
validity---and hence the success---of the formula goes well beyond
the independent particle assumption made in deriving it, and one is
left wondering about the true meaning of the parameters. The Shell
Model Monte Carlo method provides an alternative approach to level
densities~\cite{dea95,nak97,lan98} whose reliability is now
established~\cite{alh99}. The problem is that the
calculations are hard.

To combine the simplicity of Eq.~(\ref{bet}) with a parametrization of
clear microscopic origin, we propose a continuous binomial form to
describe the shell model level densities. It will be shown to do
well with experimental data, to reproduce strikingly Eq.~(\ref{bet})
over a wide range of energy, and to provide a strong form of the
central limit theorem (CLT), generalizing the famous result of Mon and
French~\cite{mon75} (MF) for Hamiltonians of arbitrary rank.

Consider a system of $m$ particles moving in $D$ orbits, spanning a
space of dimensionality $d$. To specify a  binomial density
$\rho_b(x,N,p,S)$, three parameters are needed: $N$, the effective
number of particles, the asymmetry $p$, and an energy scale
$\varepsilon$. The span (distance between lowest and highest
eigenstates), centroid $E_c$, variance $\sigma^2$ and the adimensional
energy variable $x$ are given by
\begin{equation}
  \label{npe}
  S=N\varepsilon, \quad E_c=Np\varepsilon , \quad
  \sigma^2=Npq\varepsilon^2, \quad x=\frac{E}{S},
\end{equation}

Where $p+q=1$. Calling  $\bar x=1-x$, the density is  

\begin{equation}
  \label{rbin}
\rho_b(x,N,p,S)=p^{xN}q^{\bar
  xN}d\frac{\Gamma(N+1)}{\Gamma(xN+1)\Gamma(\bar xN+1)}\frac{N}{S},  
\end{equation}
which reduces to a discrete binomial, ${N\choose n}$ if
$x=n/N=n\varepsilon/S$, with integer $n$.
 
It is often convenient to
introduce $\lambda=p/q$. Then
\begin{equation}
  \label{d0}
p^{xN}q^{\bar
  xN}d=\frac{\lambda^{xN}}{(1+\lambda)^N}d\equiv\lambda^{xN}d_0
\therefore d=d_0(1+\lambda)^N,   
\end{equation}
where $d_0$ is the number of states at $x=0$. To determine $N$, $p$
and $S$ (or $\varepsilon$), Eqs.~(\ref{npe}) for $S$ and $E_c$ cannot
be used , since the spectrum is not known. We have to rely instead on
the moments of the Hamiltonian ${\cal H}$, i. e., averages given by
the traces of ${\cal H}^K$, to be equated with the corresponding
moments of $\rho_b(x,N,p,S)$, which for low $K$ are the same as those
of the discrete binomial (Eqs.~(\ref{dom})). The necessary definitions
and equalities follow.

\begin{eqnarray}
  \label{moms}
&&  d^{-1}\text{tr}({\cal H}^K)=\langle{\cal H}^K\rangle, \;
  E_c=\langle{\cal H}^1\rangle,\; {\cal M}_K=\langle({\cal
  H}-E_c)^K\rangle\nonumber \\
&&\sigma^2={\cal M}^2,\quad \overline{\cal M}_K=\frac{{\cal
  M}_K}{\sigma^K},\quad \gamma_1=\overline{\cal
  M}_3=\frac{q-p}{\sqrt{Npq}} \nonumber \\
&&\gamma_2=\overline{\cal M}_4-3=\frac{1-6pq}{Npq}; \qquad d=d_0(1+p/q)^N.
\end{eqnarray}

$N$ and $p$ can be extracted either through the equations for
$\gamma_1$ and $\gamma_2$, or through those for $\gamma_1$ and $d$.
The former option is unambiguous, and it has the advantage of warning
us that the binomial form is doomed if $\gamma_2/\gamma_1^2>1$.  The
latter (which we adopt here) is simpler, and physically cogent for the
natural choice $d_0$=1, which locates the ground state at $x=0$. Once
$N$ and $p$ are known, $S=(N\sigma^2/pq)^{1/2}$ follows.  The centroid
$E_c$ provides the energy reference. With the simple choice, the
predicted ground state is at $E_0=-Sp$, which may not coincide with
the exact value, usually taken as origin. Therefore a shift $\Delta$
may be necessary, as in Eq.~(\ref{bet}).
 
Let us apply these prescriptions to $^{55}$Mn, $^{56}$Fe and
$^{60}$Ni, for which data are available~\cite{hui69}. There is also a
Monte Carlo calculation that does very well for the first of these
nuclei~\cite{nak97}. The space chosen by Nakada and Alhassid is the
$pf+g_{9/2}$ shells. The dimensionalities are given by $d={D\choose
  z}{D\choose n}$ where $D=30$ and $z$, $n$ are the numbers of active
protons and neutrons (e. g. $z=8$, $n=12$ for $^{60}$Ni). In general,
to proceed, we would have to calculate the three lowest moments of the
Hamiltonian, a feasible task even in enormous spaces.  Here we
simplify matters by making first the neutral choice $p=0.5$, $N=\ln
d/\ln 2=$ (42,44,49) for the three nuclei in the order above.  Then,
we bypass the calculation of $\sigma^2$ and estimate $S$ directly . A
first approximation comes from the single particle energies at
$(0,6,7,7,10)$ MeV for $(f_{7/2},p_{3/2},f_{5/2},p_{1/2},g_{9/2})$
respectively, for which the differences between highest and lowest
states come at (138,148,170) MeV for ($^{55}$Mn,$^{56}$Fe,$^{60}$Ni).
Correlations increase these numbers by an amount that ranges from 10
MeV for $^{56}$Ni to 20 MeV for $^{48}$Cr, according to exact
calculations in the $pf$ shell~\cite{cau99}. Hence the estimate
$S$=$(153\pm 5,133\pm 5,185\pm 5)$ for the trio. Fixing $N$ and
allowing $S$ to vary within the error bars, the remaining uncertainty
comes from the position of the ground state, which is also allowed to
vary within $\pm \varepsilon/2\approx 1.5$ MeV. Numbers that fall very
confortably within the assigned error bars: $S=(150,165,185),\;
\Delta=(1.15,-0.40,0.20)$ MeV, define binomial densities that agree
very well the experimental ones ~\cite{ilj92} ( where the error bars
are apparently smaller than the original ones in ~\cite{hui69}):
   \begin{figure}[htbp]
    \begin{center}
      \leavevmode
      \epsfig{file=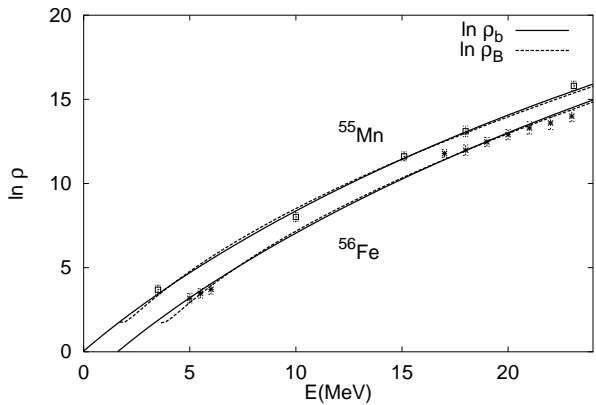,width=8cm}
   \caption{Experimental, binomial (Eq.~(\ref{rbin}))  and
      Bethe (Eq.~(\ref{bet})) logarithmic level
     densities in $^{55}$Mn and $^{56}$Fe. For parameters see text. }
      \label{fig:den55_56}
    \end{center}
  \end{figure}

  Fig.~\ref{fig:den55_56} gives the experimental points for the first
  two nuclei, the binomials (whose parameters have been estimated
  above), and fits to the binomials using Bethe's formula yielding
  $(4a,\Delta)=$ (21.5,-1.5) and (21.3,-3.4) for $^{55}$Mn and
  $^{56}$Fe respectively.

The near identity between binomial and Bethe forms
  extends to $\approx 40$ Mev. Therefore, the inmense experience
  accumulated by fitting  Eq.~(\ref{bet}) to the data can be
  reanalyzed in terms of Eq.~(\ref{rbin}) whose phenomenological
  potential has been illustrated by the simple exercise above. Its
  efficiency in dealing with rigorously calculated moments will be
  demonstrated in~\cite{zuk00}.
  
  The near identity between Bethe and binomial forms has also a
  mathematical significance that is discussed in the paragraph that
  follows the one containing Eq.~(\ref{Factor}).

\begin{figure}[htbp]
    \begin{center}
      \leavevmode
       \epsfig{file=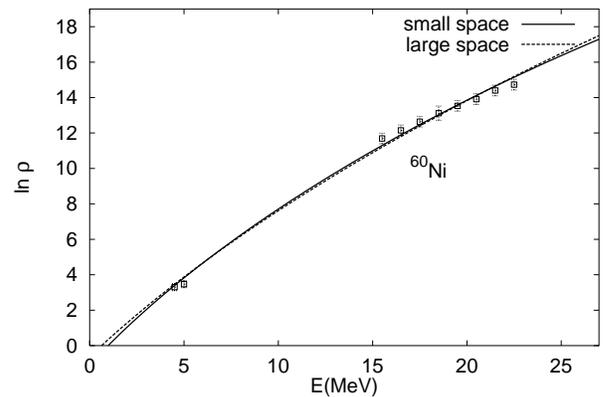,width=8cm}
\caption{Experimental logarithmic level densities for
  $^{60}$Ni, compared with binomial ones calculated in two spaces}
      \label{fig:den60}
    \end{center}
  \end{figure}

  For $^{60}$Ni in Fig.~\ref{fig:den60}, instead of showing
  $\ln\rho_B$---which fits obviously as well as in
  Fig.~\ref{fig:den55_56}---a different check is proposed. As our
  finite space can provide reliable densities only in a finite
  interval, to have an idea of its range, the space has been enlarged
  by adding the $d_{3/2}$ and $s_{1/2}$ shells, at 3 MeV below
  $f_{7/2}$. The parameters found for the smaller space
  ($N,S,\Delta$)= (49,185,0.2), become (65,285,0.8) for the larger
  one. The densities coincide nicely in the region of interest.
  Discrepancies become appreciable after 25 MeV, which provides a
  preliminary indication: binomial densities can be trusted in an
  interval of $\approx 0.15S$ above the ground state. It is very much
  in the logic of the construction, that knowledge of the level
  density up to a given energy, can be extended to higher energies.

The first part of the program is now successfully completed.

To understand in  what sense binomial and Hamiltonian moments coincide
we derive the general form for both. The following
notation will prove very useful ($a$ is an arbitrary array): 
\begin{equation}
  \label{[]}
  [a]^l_{\ne}=\sum_{\alpha_1\cdots \alpha_l}a_{\alpha_1}\cdots
  a_{\alpha_l}, \qquad \alpha_i\ne \alpha_j.
\end{equation}
If $\alpha_i$ can take $L$ values, there are $L(L-1)\cdots
(L-l+1)\equiv L^{(l)}$ terms (Boole's notation). If the indeces are
ordered, $\alpha_1<\alpha_2\cdots$, we define analogously
$[a]^l_<=[a]^l_{\ne}/l!$.

For the binomial case we start from the standard result 
\begin{eqnarray}
  \label{Mt}
{\cal M}(t)&=&  \sum_n p^nq^{N-n}{N\choose
  n}e^{t(n-Np)}=(pe^{qt}+qe^{-pt})^N
\nonumber\\
{\cal M}_K&=&
\left(\frac{\partial}{\partial t}\right)^K{\cal
  M}(t)\left|_{t=0}\right.=
{\cal  M}^{[K]}\left|_{t=0}\right.  
\end{eqnarray}
To evaluate ${\cal M}_K$, consider the multinomial expansion (Leibnitz
rule) for the $K$-th t-derivative of a product of $N$ factors
$h_{\alpha ,t}$
\begin{eqnarray}
  \label{prod}
&&  \left[\prod_{\alpha =1}^N
h_{\alpha,t}\right]^{[K]}=\sum_{\alpha,k_{\alpha}}
K!\frac{h_{1,t}^{[k_1]}}{k_1!} 
  \frac{h_{2,t}^{[k_2]}}{k_2!}\cdots=\\
&&\sum_{s,l_s}
 K!\prod\frac{\left[h^{[s]}\right]^{l_s}_<}{(s!)^{l_s}}=\sum_{s,l_s}
 K!\prod\frac{\left[h^{[s]}\right]^{l_s}_{\ne}}{(s!)^{l_s}l_s!},
\quad \sum sl_s=K.\nonumber
\end{eqnarray}
In the second equality, the $l_s$ factors with common $k_{\alpha}=s$
have been regrouped in $\left[h^{[s]}\right]^{l_s}_<$ and then the
original ordering among the factors has been relaxed. {\em Notations from
Eq.~(\ref{[]})}. Note also that (obviously) $\sum_s l_s=N$.

 Identifying $h_{\alpha,t}=pe^{qt}+qe^{-pt}$, we have
\begin{equation}
  \label{h_k}
h_{\alpha,t}^{[s]}=pq[q^{s-1}e^{qt}+(-)^sp^{s-1}e^{-pt}]. 
\end{equation}
Calling $h_s=h_{\alpha,t=0}^{[s]}$ ($h_0=1,h_1=0$), and $N-l_0=l$,
the moments follow from Eq.~(\ref{prod}), and noting that
$N^{(l)}=N!/(N-l)!=N!/l_0!$, (the $N!$ coming from the equality of all the
$h_{\alpha,t}$ factors), the result is:  
\begin{equation}
  \label{M_k}
  {\cal M}_K=K!\sum_{s,l_s}N^{(l)}\prod\frac{(h_s)^{l_s}}{(s!)^{l_s}l_s!},
\quad s\geq 2. 
\end{equation}
 The dominant terms  maximize $l$, subject to the conditions
$\sum sl_s=K$, $\sum l_s=l\leq N$. Keeping contributions up to
$O(1/N)$, the normalized moments (${\cal M_K}/{\sigma}^K$,
$\sigma^2=Nh_2$) for $K=2k$ and $K=2k+1$ respectively, are 
\begin{eqnarray}
  \label{dom}
&&  \frac{N^{(k)}}{N^k}(2k-1)!!
\left[1+\frac{1}{(N-k+1)}\left(\frac{k^{(2)}h_4}{6h_2^2}+\frac{k^{(3)}h_3^2}{9h_2^3}\right)\right],
\nonumber  \\
&& \frac{N^{(k)}}{N^k}(2k+1)!!\frac{kh_3}{3\sqrt{N}h_2^{3/2}}.
\end{eqnarray}
They will remain dominant as long as $k< \sqrt{N}$. 

To calculate moments of a Hamiltonian, we write ${\cal
  H}^K=(\sum_{\alpha}h_{\alpha,t})^K$. In the single particle case
$h_{\alpha,t}=m_{\alpha}\varepsilon_{\alpha}$, so the
  summands commute, and the multinomial expansion is {\bf exactly} the
  same as in Eq.~(\ref{prod}), with derivatives, $[s]$, replaced by
  powers $s$, and we have now
  $[h^s]^{l_s}_{\ne}=[\varepsilon^s m^s]^{l_s}_{\ne}$. 
Since $m^s_{\alpha}=m_{\alpha}$, and there are $l$ factors, the number
operators contribute as $\prod_{i=1,l}m_{\alpha_i}$, 
whose trace is $m^{(l)}/D^{(l)}$ (it must be proportional to
$m^{(l)}$, and unity at the closed shell in which the $D$ orbits are
  full). Defining 
  $\epsilon^s_\alpha=\varepsilon^s_\alpha D^{-1}$
  the moments for a one body ($r=1$) Hamiltonian follow as
\begin{equation}
  \label{M_k1}
  {\cal M}_K^{r=1}=K!\sum_{s,l_s}m^{(l)}
\frac{D^l}{D^{(l)}}\prod\frac{[\epsilon^s]^{l_s}_{\ne} }{(s!)^{l_s}l_s!}
\quad s\geq 1. 
\end{equation}
In the dilute limit ($m\ll D$), $D^l/D^{(l)}\approx 1$, and
$[\epsilon^s]^{l_s}_{\ne}\approx
(\sum_{\alpha}\epsilon^s_{\alpha})^{l_s}\equiv \langle \epsilon^s
\rangle ^{l_s}$, the notation in MF, whose
result~\cite[Eq.~(7)]{mon75} is identical to Eq.~(\ref{M_k1}) (the
$s=1$ terms cancel since ${\cal H}$ is taken to be traceless). The
complete analogy with Eq.~(\ref{M_k}) is obvious.

To understand how to dispose of the condition  $m\ll D$, let us
calculate $\sigma^2(m)$, (use $\bar{m}=D-m$)
\begin{eqnarray}
\label{s^2}
&&\langle \sum_{\alpha}\varepsilon_{\alpha}^2m_{\alpha}+
\sum_{\alpha \ne \beta}\varepsilon_{\alpha}\varepsilon_{\beta}
m_{\alpha}m_{\beta}\rangle_m=\nonumber\\
&&\sum_{\alpha}\varepsilon_{\alpha}^2\left[\frac{m}{D}
-\frac{m^{(2)}}{D^{(2)}} \right]
=\frac{m\bar{m}}{D-1}\sum_{\alpha}\frac{\varepsilon_{\alpha}^2}{D},
\end{eqnarray}
where we have added and subtracted the terms necessary to eliminate
the restriction $\alpha \ne \beta$ and remembered that
$\sum_{\alpha}\varepsilon_{\alpha}=0$. Similarly, for the third moment
we find 
\begin{equation}
  \label{m_3}
{\cal M}_3^{r=1}=\frac{m\bar{m}(\bar{m}-m)}{(D-1)^{(2)}}
\sum_{\alpha}\frac{\varepsilon_{\alpha}^3}{D}.    
\end{equation}
Under particle-hole transformation (i. e. $m\longrightarrow \bar{m}$),
the even moments are symmetric, and the odd ones antisymmetric, which
simplifies enormously the calculations. E. g., Eq.~(\ref{s^2}) follows
from the argument that $\sigma^2(m)$ must be a two body operator that
reduces to the correct value for $m, \bar{m}=1$. More generally, the
leading term for $\overline {\cal M}_{2k}^{r=1}={\cal
  M}_{2k}^{r=1}/(\sigma^2(m))^k$ must have the form
\begin{equation}
  \label{Factor}
  \frac{m^{(k)}\bar{m}^{(k)}[D^{(2)}]^k}{m^k\bar{m}^kD^{(2k)}}
(2k-1)!!={\cal F}(m,k)(2k-1)!!,
\end{equation}
which follows from demanding rank $2k$, symmetry in particles and
holes, vanishing for $m, \bar{m} <k$ and corrrect value at $m,
\bar{m}=k$ (i. e. $d^{-1}_k=k!/D^{(k)}$). The important point is that
${\cal F}(m,k)\approx 1$ as long as $k\ll m$, i. e., {\em as long as
  this term remains dominant the moments are those of a Gaussian}
(CLT). When the leading odd moments and subdominant even ones are
included we end up with an expression identical to Eq.~(\ref{dom})
once $N$ and $p$ have been extracted as explained at the beginning.
Hence, we have proved that the level density for a one body
Hamiltonian has binomial moments for $K<\sqrt N$. A strong version of
the central limit theorem (CLT is a statement about the even-$K$
dominant term only).

The near identity between Eqs.~(\ref{bet}) and~(\ref{rbin}) is an
excellent reason to expect that for single particle 
Hamiltonians the binomial behaviour extends to the full spectrum,
since Eq.~(\ref{bet}) is a rigorous mathematical result that applies
in this case. 

For a Hamiltonian of higher rank, ${\cal H}=\sum_{xy}W_{xy}Z^+_xZ_y$,
where $Z^+_x$, $Z_y$ create and annihilate $r$ particles, the
operators do not commute. Nevertheless, Eq.~(\ref{dom}) is of use in
giving the correct counting: there are $(2k-1)!!$ ways of contracting
${\cal H}^{2k}$ in $k$ pairs, $k(2k+1)!!/3$ ways of contracting ${\cal
  H}^{2k+1}$ in $k-1$ pairs and one triple, etc. However, the
contributions of each term are different. In other words: for the
dominant term, say, the factor ${\cal F}(m,k)$ becomes extremely
complicated. The problem was solved by Mon and French~\cite{mon75}.
Here we give an idea of their result.
For simplicity assume that ${\cal H}$ is a two-body operator ($r$=2),
and stay in the dilute limit. Then the variance 
$\sigma^2(m)=\langle {\cal H}^{2} \rangle_m/d_m$ must be 
 \begin{equation}
\sigma^2(m)=\frac{m^{(2)}}{2}\sigma^2(2)\,,\quad  \sigma^2(2)=
\frac{\sum W_{xy}^2}{d_2},
\label{eq:m2}
\end{equation}
Now consider $\langle {\cal H}^{4} \rangle$, leading to three
possible contractions, written in MF as ${\cal H}{\cal H}{\cal H}{\cal
  H}=AABB+ABBA+ABAB$. The first two give
$m^{(2)}\sigma^2(2))^2/4=(\sigma^2(m))^2$, but for the last one we
have $m^{(4)}(\sigma^2(2))^2/4$, which vanishes at $m=2$. The general
result is that for $m=2$ and $K=2k$, only \mbox{$t_k={2k \choose k -1}/k$}
terms survive, the Catalan numbers, i. e., the normalized moments of
the semicircular density for Wigner's GOE ensemble. As $m$ increases
the number of surviving terms increases rapidly so as to have again
${\cal F}(m,k)\approx 1$.

{\em It is straightforward to apply similar arguments to the other dominant
and subdominant terms}. Thus, we can combine the Mon French analysis and
 the advantages of the binomial geometry, to obtain a strong
form of the CLT, seen to apply generally to higher rank
Hamiltonians.

So far the good news. There are no bad news, but a hard question: Why
stop the analysis at the low moments? As mentioned, it is practically
certain that for rank one the proof of binomial behaviour must extend
to the full spectrum. The strong formal analogy between rank one and
higher should encourage the generalization. The catch is that, more
often than not, systems undergo phase transitions. As there is little
risk in attributing them to some form of collectivity, we can guess
that the binomial forms will be valid in the absence of strong enough
collectivity. Random Hamiltonians fulfill this condition, and I
propose to include them in a larger class: {\em A Hamiltonian {\em
    acts as random} if it does not have strong enough collective
  components}.

There may be two reasons for the good performance of binomials in the
nuclear case. One is the strong dominance of the single particle
field. The other is the lack of sufficient collectivity in the nuclear
Hamiltonian. It is certainly not random, as it contains sizeable
pairing and quadrupole forces~\cite{dz96}. However, they do not seem
to be strong enough. A quantitative estimate of the relative strength
of the different components is given by $\sigma^2$: we have already
seen that the single particle contribution is far stronger that the
two body part, and from~\cite{dz96} we know that in the latter,
pairing plus quadrupole only contribute a fraction of the total.
Sufficient to give them a capital spectroscopic role but, apparently
not the possibility of distorting the binomial forms.

The formalism is ready to examine the problem, which will become even
more interesting with the suggestion that Hamiltonian matrices at
fixed quantum numbers always have binomial level densities~\cite{zuk00}.

\begin{acknowledgements}
  
  Professor J. B. French provided some very early encouragements. G.
  Dussel insisted on the use of continuous binomials, and G.
  Mart\'{\i}nez Pinedo on the use of exact continuous binomials. I had
  some very useful discussions with Y. Alhassid, V. K. Kota and
  A.Poves.

This work is dedicated to Emilia Montaner.
\end{acknowledgements}

\end{multicols}
\end{document}